\documentclass[aps,prd,twocolumn,superscriptaddress]{revtex4-1}
\usepackage{color}
\usepackage{soul}
\usepackage{graphicx}
\usepackage{dcolumn}

\begin{document}


\title{Characterization of positronium properties in doped liquid scintillators}


\author{G.~Consolati}
\affiliation{Department of Aerospace Science and Technology, Politecnico di Milano, via La Masa 34, 20156 Milano, Italy}

\author{D.~Franco}
\affiliation{APC, AstroParticule et Cosmologie, Universit\'e Paris Diderot, CNRS/IN2P3, 75205 Paris, France}
\author{S.~Hans}
\affiliation{Departement of Chemistry, Brookhaven National Laboratory, Upton, NY}
\author{C.~Jollet}
\affiliation{IPHC, Universit\'e de Strasbourg, CNRS/IN2P3, 67037 Strasbourg, France}
\author{A.~Meregaglia}
\affiliation{IPHC, Universit\'e de Strasbourg, CNRS/IN2P3, 67037 Strasbourg, France}
\author{S.~Perasso}
\affiliation{APC, AstroParticule et Cosmologie, Universit\'e Paris Diderot, CNRS/IN2P3, 75205 Paris, France}
\author{A.~Tonazzo}
\affiliation{APC, AstroParticule et Cosmologie, Universit\'e Paris Diderot, CNRS/IN2P3, 75205 Paris, France}
\author{M.~Yeh}
\affiliation{Departement of Chemistry, Brookhaven National Laboratory, Upton, NY}


\begin{abstract}


 Ortho--positronium (o-Ps) formation and decay can replace  the annihilation process, when positron interacts in liquid scintillator media. The delay induced by the positronium decay represents either a potential signature for anti--neutrino detection, via inverse beta decay, or to identify and suppress positron background, as recently demonstrated by the Borexino experiment.\\ 
The formation probability and decay time of o-Ps depend strongly on the surrounding material. In this paper, we characterize the o-Ps properties in liquid scintillators as function of concentrations of gadolinium, lithium, neodymium,  and tellurium, dopers used by present and future neutrino experiments. In particular, gadolinium and lithium are high neutron cross section isotopes, widely used in reactor anti--neutrino experiments, while neodymium and tellurium are double beta decay emitters,  employed  to investigates the Majorana neutrino nature. Future neutrino experiments may profit from the performed measurements to tune the preparation of the scintillator in order to maximize the o-Ps signature, and therefore the discrimination power.

\end{abstract}


\maketitle

\section{Introduction}

Particle detection in liquid scintillators benefits of a powerful technique to discriminate among interacting particles: the light pulse shape discrimination (PSD), a key feature to extract the searched signal from the background.\\
The PSD relies on the different time profiles of the scintillation photon emission, dependent on the energy loss and hence on the kind of ionizing particle crossing the media.
This technique is particularly effective in discriminating light particles, as electrons and positrons, from heavy particles, like protons, ions and alpha particles (see Ref.~\cite{Ranucci:1998bc} and references therein). However, such  technique  is inadequate to distinguish between electrons and positrons.

Recently, a new PSD technique \cite{Franco:2010rs} has been proposed to separate electrons from positrons. 
The latter, in fact, may be identified by exploiting the formation of a metastable electron--positron bound state (positronium), a competitive process with respect to direct annihilation.
The  positron--induced pulse shape in liquid scintillators is indeed the sum of two components: the positron ionization and the positronium decay $\gamma$'s. The latter is delayed by the positronium mean lifetime.\\ 
Positronium exists in two states: the singlet, called para--positronium  (p-Ps), which decays into two $\gamma$'s with a lifetime in vacuum of 125~ps, and the triplet, or ortho--positronium (o-Ps), which decays into three $\gamma$'s with a lifetime of 142~ns. Triplet and singlet states are formed in a ratio of 3:1.\\
In the case of p-Ps, the delay between the ionization and the annihilation $\gamma$'s is negligible with respect to the characteristic detection times.\\
In matter, o-Ps is subjected to chemical reactions (oxidation or compound formation), magnetic effects (spin--flip), and interactions with the surrounding electrons (pick--off, the dominant effect in liquid scintillator in absence of electric and magnetic fields).
These interactions yield a two body decay~\cite{PsBook} and cause a sizable shortening of the o-Ps lifetime to a few nanoseconds~\cite{Franco:2010rs}.\\
Nevertheless, even such an o-Ps mean lifetime may induce an observable distortion in the photon emission time distribution.

The signature provided by the o-Ps--induced pulse shape distortion has already been successfully exploited by the Borexino collaboration \cite{Collaboration:2011nga} in the identification and rejection of the cosmogenic $^{11}$C $\beta^+$ decays, the dominant  background in the solar {\it pep} neutrino rate measurement.\\
More in general, cosmic muon interactions in organic liquid scintillators produce several other sources of $\beta^+$ decays, such as $^8$B, $^9$Be, and $^{10}$C \cite{Collaboration:2010b8}, which represent critical contaminations in underground low-background neutrino 
experiments, such as Borexino~\cite{Collaboration:2009det} and SNO+~\cite{Chen:2005}.\\
Furthermore, the o-Ps--enhanced PSD may strengthen the electron anti--neutrino detection, usually performed via the inverse beta decay process: $\bar{\nu}_e + p \to e^+ + n$. 
The positron identification, in addition to the positron--neutron  delayed coincidence, can abate the rate of random coincidences and correlated electron--neutron background, such as the cosmogenic $^9$Li and $^8$He decays. 
This technique can be applied in reactor neutrino experiments, like Double Chooz~\cite{Ardellier:2006mn}, Daya Bay~\cite{Guo:2007ug}  and RENO~\cite{Ahn:2010vy}.

The o-Ps properties (lifetime and formation probability) have already been measured in the solvents for organic liquid scintillators commonly used in neutrino physics (pseudocumene (PC), linear alkylbenzene (LAB), phenylxylylethane (PXE), and dodecane) and in scintillator mixtures, based on pseudocumene and 2,5-diphenyloxazole (PPO) with or without isoparaffin \cite{Franco:2010rs, Kino:2000}.\\
Recent progress in chemistry allows to obtain stable scintillators loaded with organo-metallic compounds. 
Elements like gadolinium 
 and lithium are employed to enhance neutron detection, thanks to their large cross section. 
 In particular, gadolinium is commonly used  to improve the signature  in reactor anti--neutrino experiments, 
whereas lithium is particularly useful in neutron detection due to charged particle production after capture (see e.g. Ref.~\cite{Fisher:2011hm}). 
Both metal-doped scintillators could also be used as a veto for low background experiments such as direct dark matter search ones.\\
Scintillators can also be loaded directly with the signal source, like in the case of SNO+, where the double beta emitter, originally  $^{150}$Nd, and very recently replaced by $^{130}$Te~\cite{SNOLi}, is mixed with the active mass.

In this work, we measured the effects of the gadolinium, neodymium, lithium and tellurium compounds on the positronium properties in liquid scintillator, as a function of their concentrations.

\section{Experimental Setup}
A standard PALS system made of two plastic scintillators (Pilot U) detectors was used to measure the o-Ps formation fraction and its lifetime. 
A detailed description of the apparatus can be found in Ref.~\cite{Franco:2010rs}. 
The $^{22}$Na positron source, deposited between two Kapton${\textregistered}$ (DuPont) layers 15~$\mu$m thick, is immersed in the vial containing scintillator to be tested. 
The first plastic scintillator is configured (lower energy threshold at 900~keV) to produce a trigger signal detecting the 1.27~MeV $\gamma$ emitted in the $^{22}$Na decay along with a positron.
The other detector (350 to 500~keV energy range) generates a second signal when a 511 keV from the positron annihilation is revealed.
The difference in time between the two signals is measured to reconstruct the o-Ps lifetime.

\begin{figure}[htbp]
\begin{center}
\includegraphics[width=9cm]{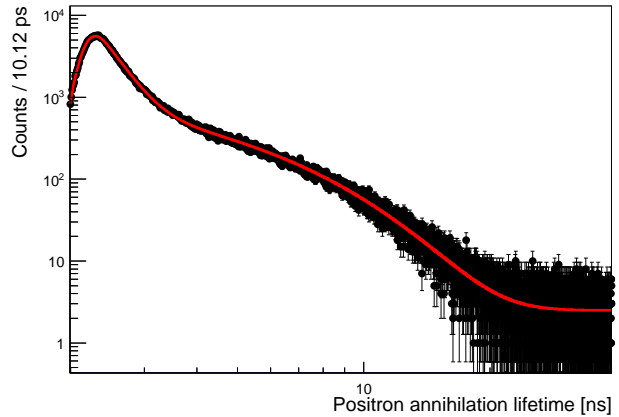}
\caption{Positron annihilation lifetime measured in the LAB sample with no doper (black dots). The fit function is shown in red.}
\label{fig:spectra}
\end{center}
\end{figure}

A typical observed spectrum can be seen in Fig.~\ref{fig:spectra}. 
The fit is performed using the RooFit package~\cite{Verkerke:2003ir} based on MINUIT. 
The fit function is a combination of three exponentials and a constant term:
\begin{equation}
F(t)= \sum\limits_{i=1}^3 A_i\cdot e^{-t/\tau_i} + C 
\end{equation}
where $A_i$ and $\tau_i$  (i = 1, 2) correspond to the effective amplitude and lifetime of direct annihilation and p-Ps; A$_3$ and $\tau_3$ refer instead to o-Ps.
The constant term $C$ accounts for the accidental background.
The use of two exponentials for the description of direct annihilation and p-Ps is a standard practice in positron annihilation spectroscopy, and depends on the fact that the positron thermalization lifetime is different in the source support (Kapton in our case) and in the medium to be tested (liquid scintillator). 

The fit function $F(t)$ is convoluted with a gaussian distribution to model the detector resolution ($\sigma~\sim$~120~ps).\\

\section{Data Analysis}

The o-Ps properties are extracted from the fit result.
The fit parameter $\tau_3$ is a direct measurement of the o-Ps lifetime.
It consists in an effective value resulting from the two and three gamma decay modes, whose fractions are respectively:
\begin{equation}
f_{2\gamma}=1-\frac{\tau_3}{\tau_{3\gamma}} \;\;\;  \mathrm{and} \;\;\; f_{3\gamma}=1-f_{2\gamma}
\end{equation}
where $\tau_{3\gamma}$ is the vacuum o-Ps lifetime of 142~ns.

The evaluation of the o-Ps probability formation ($f$) is less straightforward.
Assuming a different detection efficiency for the two ($\epsilon_2$) and three gamma ($\epsilon_3$) decays, the number of annihilation ($A_A = A_1 + A_2$) and o-Ps formation ($A_3$) measured are:
\begin{equation}
A_A = A_K + (1-f) C_S \epsilon_2
\label{eq:A1}
\end{equation}
\begin{equation}
A_3 =  f C_S (f_{2\gamma} \epsilon_2 + f_{3\gamma} \epsilon_3)
\label{eq:A2}
\end{equation}
where $A_K$ is the number of annihilations measured in the Kapton and $C_S$ the total number of events in the scintillator volume.\\
The fraction of annihilations in Kapton, $A_K$, was extrapolated from measurements at various Kapton layer thicknesses and found to be 
$20.6 \pm 0.2\%$~\cite{Franco:2010rs}.\\
Solving the system of Eq.~\ref{eq:A1} and \ref{eq:A2}, the o-Ps probability formation is found to be:
\begin{equation}
f = \frac{A_3 \tau_{3\gamma}} { (A_A + A_3 - A_K) \tau_3 + (A_A -A_K) (\frac{\epsilon_3}{\epsilon_2} -1) \tau_3}.
\label{eq:f}
 \end{equation}
 
As it can be seen in Eq.~\ref{eq:f}, the evaluation of the o-Ps formation probability relies on the knowledge of the ratio of the detection efficiencies for the three and two gamma channels.\\
Since this value could not be measured with high precision in the experiment, the o-Ps formation fraction was computed in the two extreme cases $\epsilon_3 = 0$ and $\epsilon_3 = \epsilon_2$.
The average of the two obtained values was taken as measure of $f$, while the difference was taken as contribution to the systematic error (1.2\%).
Furthermore, another component of the systematic error was estimated looking at the discrepancies between the measurements of the same sample (each one was measured three times): this resulted in an error of about 1.3\%.\\
The error budget on the o-Ps formation fraction includes also a statistical component, given by the error propagation in Eq.~\ref{eq:f}, which is typically of the order of 0.6\%.
The same was done for the lifetime evaluation where an error of about 0.9\% (corresponding to $\sim 0.03$~ns) was found. 
Moreover, an error of 0.3\% on the time coming from the setup calibration procedure is included.

To summarize, adding each error contribution quadratically the error on the o-Ps formation fraction obtained is $\sim  1.9\%$ whereas $\sim  1\%$ error is obtained on its lifetime.

\section{Results}

\subsection{LAB+Gd}
\label{sec:Gd}
The Gd doped sample is a LAB based scintillator mixed with 3~g/L of PPO, 15~mg/L of 1,4-Bis(2-methylstyryl)benzene (bis--MSB)  and Gd at a concentration varying from 0.01\% to 0.45\%.\\ 
The obtained scintillator, under minor modifications, is the one typically used in reactor antineutrino experiments, such as 
Daya Bay~\cite{Guo:2007ug} and RENO~\cite{Ahn:2010vy}.

As it can be seen in
Tab.~\ref{tab:ResGd}, the o-Ps formation fraction decreases as the Gd concentration increases, whereas the lifetime is almost constant. 
It can be noted that when a very small fraction of doper is added (0.01\%) the formation fraction increases slightly with respect to the case of pure LAB: this is a known effect as explained in Ref~\cite{Djermouni}. 

%
%

\begin{table}[htdp]
\begin{center}
\begin{tabular}{ccc}
Gd concentration & $f$ & $\tau_3$\\
$[\%]$ & &$[$ns$]$\\
\hline
0 & 0.544 $\pm$ 0.008&3.05 $\pm$ 0.03\\
0.01 & 0.554 $\pm$ 0.008& 3.07 $\pm$ 0.03\\
0.05 &0.540 $\pm$ 0.008&3.05 $\pm$ 0.03\\
0.08 &0.537 $\pm$ 0.008&3.04 $\pm$ 0.03\\
0.1 &0.529 $\pm$ 0.008&3.09 $\pm$ 0.03\\
0.45 &0.406 $\pm$ 0.008&3.02 $\pm$ 0.03\\
\hline
\end{tabular}
\caption{Results for the o-Ps formation fraction and mean lifetime in Gd doped LAB samples.}
\label{tab:ResGd}

\end{center}
\end{table}%

The trends of the o-Ps lifetime and formation fraction as a function of the doper concentration are shown in Fig.~\ref{fig:time} and \ref{fig:frac} respectively, directly compared with the ones obtained using Nd as a doper (see next section). \\

\subsection{LAB+Nd}
\label{sec:Nd}
The Nd doped sample is a LAB scintillator mixed with 2~g/L of PPO and a concentration of Nd ranging from 0.05\% to 0.5\%.\\
Until the very recent proposal to use Tellurium, such a scintillator has been for a long time the best candidate for the SNO+ experiment~\cite{Chen:2005} in the search of the $0\nu\beta\beta$ decay.

Similarly to the case of the Gd loaded scintillator, the o-Ps formation fraction decreases as the Nd concentration increases, whereas the lifetime is almost constant (see Tab.~\ref{tab:ResNd}).

%
%

\begin{table}[htdp]
\begin{center}
\begin{tabular}{ccc}
Nd concentration & $f$ & $\tau_3$\\
$[\%]$ & &$[$ns$]$\\
\hline
0 & 0.537 $\pm$ 0.013&3.15 $\pm$ 0.04\\
0.05 & 0.527 $\pm$ 0.013& 3.11 $\pm$ 0.04\\
0.1 &0.494 $\pm$ 0.013&3.17 $\pm$ 0.04\\
0.3 &0.460 $\pm$ 0.013&3.15 $\pm$ 0.04\\
0.5 &0.402 $\pm$ 0.013&3.15 $\pm$ 0.04\\
\hline
\end{tabular}
\caption{Results for the o-Ps formation fraction and mean lifetime in Nd doped LAB samples.}
\label{tab:ResNd}

\end{center}
\end{table}%

\begin{figure}[ht]
\begin{center}
\includegraphics[width=9cm]{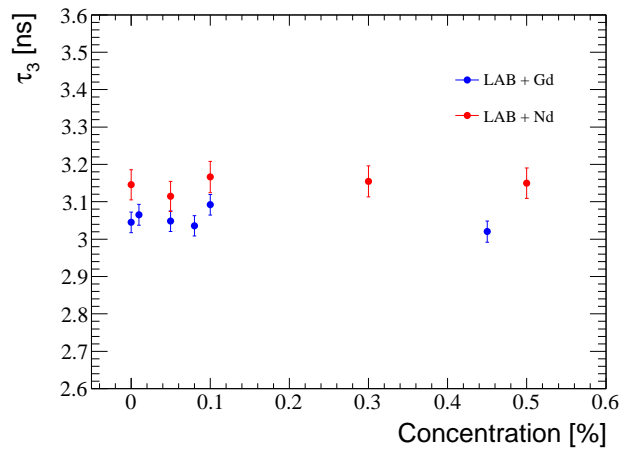}
\caption{Results for the o-Ps lifetime for Gd (blue) and Nd (red) dopers in LAB as a function of the doper concentration.}
\label{fig:time}
\end{center}
\end{figure}

\begin{figure}[ht]
\begin{center}
\includegraphics[width=9cm]{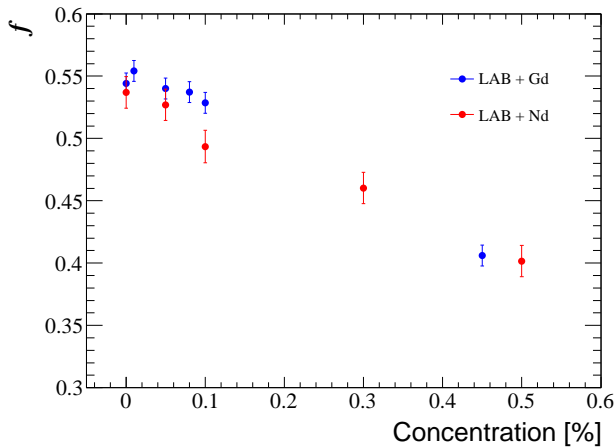}
\caption{Results for the o-Ps formation fraction for Gd (blue) and Nd (red) dopers in LAB as a function of the doper concentration.}
\label{fig:frac}
\end{center}
\end{figure}

As it can be seen in Fig.~\ref{fig:time} and \ref{fig:frac},  
o-Ps has a slightly shorter lifetime ($\sim 3\%$) in the Gd loaded scintillator than in Nd loaded one. 
This could depend on the different PPO concentration, although a previous work indicates a longer o-Ps lifetime at higher PPO concentration~\cite{Franco:2010rs}, or on the presence of bis--MSB in the Gd loaded sample.

\subsection{LAB+Li}
A different technique with respect to Gd and Nd is needed to load LAB with Li in a stable way.
Li has to be used in water solution due to its hydrophilic nature, with the net result of a water fraction in the final scintillator sample. 
In addition, the surfactant, a OH function group that mixes the water and the LAB together, accounts for about 29\% of the fractional mass.\\
The final mixture contains also 3~g/L of PPO and 15~mg/L of bis--MSB. 
 
As the fraction of water in the sample grows with the Li concentration, each sample was tested with and without Li in order to disentangle the effect of water on o-Ps from the effect of Li.\\
It can be seen in Tab.~\ref{tab:ResLi} that the lifetime is almost constant and unaffected by both Li and water. 
On the other hand, the o-Ps formation fraction is strongly affected by the presence of surfactant: even with the smallest concentration of water (0.24\%) and no Li, it is at a level of 0.363, to be compared to about 0.54 in pure LAB for Gd and Nd loaded samples (see Sec.~\ref{sec:Gd} and \ref{sec:Nd}). 

\begin{table}[htdp]
\begin{center}
\begin{tabular}{ccccc}
\multicolumn{3}{c}{concentration of}  &  &  \\
  Li  & water  & surfactant& $f$ & $\tau_3$  \\
$[\%]$ &$[\%]$ &$[\%]$ & &$[$ns$]$\\
\hline
0.01  & 0.24 & 29.93 &0.363 $\pm$ 0.011 & 2.92 $\pm$ 0.04\\
0.05  & 0.97 & 29.68 &0.353 $\pm$ 0.010 & 2.84 $\pm$ 0.03\\
0.1  & 1.99 & 29.39 &0.346$\pm$ 0.011 & 2.90 $\pm$ 0.03\\
0.35  & 6.7  & 28.02 &0.323 $\pm$ 0.010&2.90 $\pm$ 0.03\\
\hline
0& 0.24 & 29.93 &0.380 $\pm$ 0.011 & 2.92 $\pm$ 0.04\\
0 & 0.97 & 29.71 &0.367 $\pm$ 0.010 & 2.84 $\pm$ 0.03\\
0& 1.99 & 29.42 &0.351 $\pm$ 0.011 & 2.90 $\pm$ 0.03\\
0& 6.7  & 28.12 &0.344 $\pm$ 0.010&2.90 $\pm$ 0.03\\
\hline
\end{tabular}
\caption{Results for the o-Ps formation fraction and mean lifetime in Li doped LAB samples, and in the same samples without Li (same water concentration).}
\label{tab:ResLi}

\end{center}
\end{table}%

\begin{figure}[ht]
\begin{center}
\includegraphics[width=9cm]{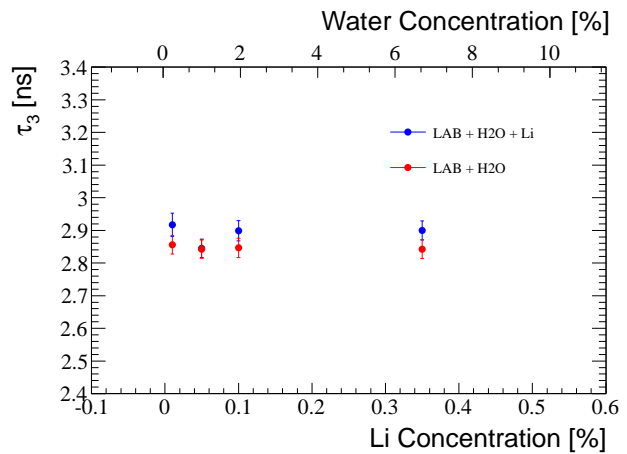}
\caption{Results for the o-Ps lifetime for Li in LAB as a function of the doper and water concentration.}
\label{fig:timeLi}
\end{center}
\end{figure}

\begin{figure}[ht]
\begin{center}
\includegraphics[width=9cm]{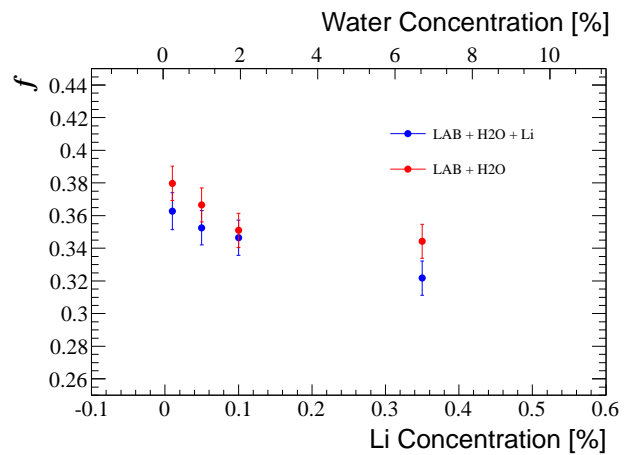}
\caption{Results for the o-Ps formation fraction for Li  in LAB as a function of the doper and water concentration.}
\label{fig:fracLi}
\end{center}
\end{figure}

The trends of the o-Ps lifetime and formation fraction as a function of the Li and water concentrations are shown in Fig.~\ref{fig:timeLi} and \ref{fig:fracLi} respectively. \\
The o-Ps formation fraction 
shows a trend similar to that of Gd and Nd doped scintillators, with the probability decreasing with increasing doper/water concentration. 
However, the impact of Li is rather weak since we obtained a difference with respect to the same sample with no Li larger than the errors only at the highest tested concentration (0.35\%). 
 This can be understood considering that the effect of the surfactant is dominant, making the effect of Li not significant.\\
In addition, the o-Ps formation fraction absolute reduction due to water/Li is at the level of $\sim ~2\%$ at the most (see Fig.~\ref{fig:fracLi}), whereas it is of the order of 15\% for Gd and Nd as dopers (see Fig.~\ref{fig:frac}) with the tested concentrations. 

\subsection{LAB+Te}

In order to obtain a stable Te loaded scintillator, a water solution is needed as in the case of Li. 
Therefore, even in this case the final sample contains a water fraction and a surfactant.
However, 
the water fraction is about a factor 3 less than the amount present in the Li loaded scintillator.\\
The surfactant is different, being an amine group instead of the OH functional group used for Li, and it is about a factor of 6 less than the amount present in the Li sample, accounting for $\sim 5\%$ of the fractional mass.\\ 
In addition, 2~g/L of PPO are present in the mixture.

As it can be seen in Tab.~\ref{tab:ResTe}, both the lifetime and the o-Ps formation probability seems unaffected by either Te or water.\\
As it was already observed in case of Li loaded scintillator, the strongest impact on the reduction on the positronium formation with respect to the pure LAB comes from the presence of the surfactant. Even with the smallest concentration of water (0.06\%) and no Te, we obtain a formation fraction of 0.359.

\begin{table}[htdp]
\begin{center}
\begin{tabular}{ccccc}
\multicolumn{3}{c}{concentration of}  & &  \\
  Te  & water  & surfactant&$f$ & $\tau_3$\\
$[\%]$ &$[\%]$ &$[\%]$ & &$[$ns$]$\\

\hline
0.01  & 0.06 & 5.00 &0.360 $\pm$ 0.009 & 2.67 $\pm$ 0.04\\
0.05  & 0.38 & 4.98 &0.363 $\pm$ 0.009 & 2.73 $\pm$ 0.07\\
0.1  & 0.57 & 4.97 &0.355 $\pm$ 0.009&2.67 $\pm$ 0.05\\
0.3  & 1.7  & 4.90  & 0.356 $\pm$ 0.009 & 2.69 $\pm$ 0.05\\

\hline
0 & 0.06 & 5.00 &0.359 $\pm$ 0.009&2.69 $\pm$ 0.04\\
0 & 0.38 & 4.98 &0.360 $\pm$ 0.009&2.69 $\pm$ 0.07\\
0& 0.57 & 4.97 &0.359 $\pm$ 0.009&2.74 $\pm$ 0.07\\
0& 1.7  & 4.90 & 0.366  $\pm$ 0.009 & 2.75 $\pm$ 0.05\\
\hline
\end{tabular}
\caption{Results for the o-Ps formation and mean lifetime in Te doped LAB samples, and in the same sample without Te (same water concentration).}
\label{tab:ResTe}

\end{center}
\end{table}%

\begin{figure}[ht]
\begin{center}
\includegraphics[width=9cm]{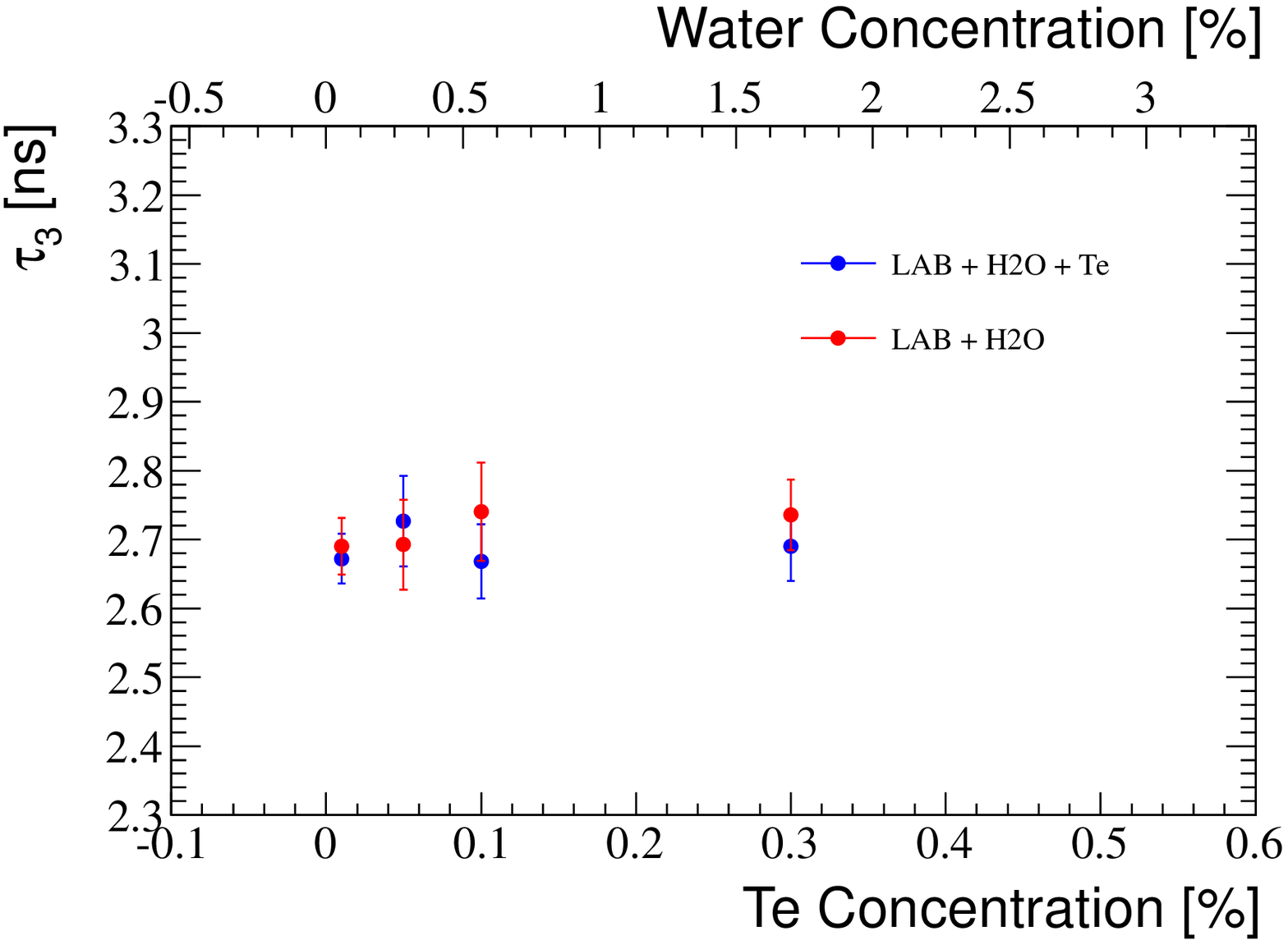}
\caption{Results for the o-Ps lifetime for Te in LAB as a function of the doper and water concentration.}
\label{fig:timeTe}
\end{center}
\end{figure}

\begin{figure}[ht]
\begin{center}
\includegraphics[width=9cm]{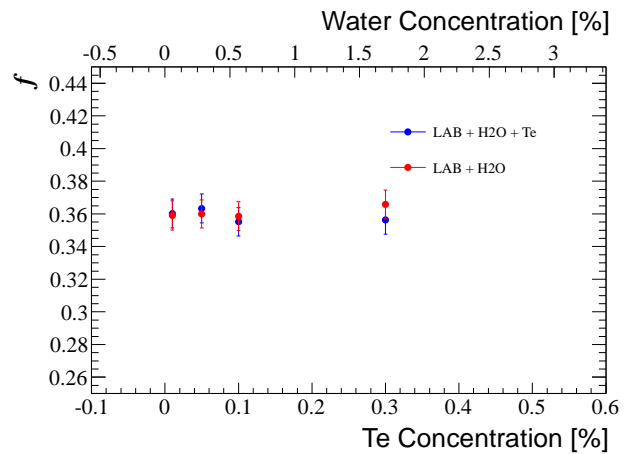}
\caption{Results for the o-Ps formation fraction for Te  in LAB as a function of the doper and water concentration.}
\label{fig:fracTe}
\end{center}
\end{figure}

The trends of the o-Ps lifetime and formation fraction as a function of the Te and water concentrations are shown in Fig.~\ref{fig:timeTe} and \ref{fig:fracTe} respectively. \\

\section{Conclusions}

The o-Ps formation fraction and lifetime were measured in different doped liquid scintillators. In particular we studied the dependence of the o-Ps properties as a function of the doper concentration, using gadolinium, neodymium, lithium and tellurium as dopers.

As first result, we observed that  the o-Ps lifetime is constant and unaffected by the doper. \\
The formation fraction is instead sensitive to the presence of a metal in the LAB and it typically decreases as the doper concentration increases. 
This behavior is particularly evident for the Gd and Nd loaded scintillators.\\
Scintillators loaded with Li and Te have a slightly different behavior due to the different loading procedure which results in the presence of water and a surfactant in the final sample. We found that both the presence of the doper and of the water have an almost negligible impact with respect to the surfactant one.

In conclusion, the o-Ps pulse shape discrimination can be exploited in both anti--neutrino and neutrino--less double beta decay experiments, to enhance the anti-neutrino signal or to reject positron like background, respectively. However, if in the first class of experiments, high concentrations of high neutron cross section dopers  can reduce by up to $\sim$25\% the o-Ps formation, the double beta emitters must be embedded in molecules containing surfactants, which suppress by up to $\sim$40\% the o-Ps component with respect to the undoped scintillator mixtures.


\section{Acknowledgments}
We acknowledge the financial support from the ANR NuToPs project (grant 2011-JS04-009-01) and from the UnivEarthS Labex program of Sorbonne Paris Cit\'e (ANR-10-LABX-0023 and ANR-11-IDEX-0005-02). The work conducted at Brookhaven National Laboratory was supported by the U.S. Department of Energy under contract DE-AC02-98CH10886.\\ We thank P.~Crivelli for useful discussions on positronium physics.


\begin{thebibliography}{99}

\bibitem{Ranucci:1998bc} 
  G.~Ranucci, A.~Goretti and P.~Lombardi,
  Nucl.\ Instrum.\ Meth.\ A {\bf 412}, (1998) 374.
  
\bibitem{Franco:2010rs} 
  D.~Franco, G.~Consolati and D.~Trezzi,
  Phys.\ Rev.\ C {\bf 83}, (2011) 015504.
  
  \bibitem{PsBook}
  H.~J.~Ache,
 http://pubs.acs.org/isbn/9780841204171.
  
\bibitem{Collaboration:2011nga} 
  G.~Bellini {\it et al.}  [Borexino Collaboration],
  Phys.\ Rev.\ Lett.\  {\bf 108}, (2012) 051302.
  
\bibitem{Collaboration:2010b8}
 G.~Bellini {\it et al.}  [Borexino Collaboration],
 Phys. Rev. D {\bf 82}, (2010) 033006.

\bibitem{Collaboration:2009det}
G.~Alimonti {\it et al.}  [Borexino Collaboration],
Nucl. Instrum. Meth. A {\bf 600}, (2009) 568-593.

\bibitem{Chen:2005}
M.~C.~Chen, Nucl. Phys. Proc. Suppl., {\bf 145} (2005) 65.
  

 %



\bibitem{Ardellier:2006mn} 
  F.~Ardellier {\it et al.}  [Double Chooz Collaboration],
  hep-ex/0606025.
%

  

  
\bibitem{Guo:2007ug} 
  X.~Guo {\it et al.}  [Daya-Bay Collaboration],
  hep-ex/0701029.
  
\bibitem{Ahn:2010vy} 
  J.~K.~Ahn {\it et al.}  [RENO Collaboration],
  arXiv:1003.1391 [hep-ex].

\bibitem{Kino:2000}
Y. Kino {\it et al.}, J. Nucl. Radiochem. Sci. {\bf 1} (2000) 63.



\bibitem{Fisher:2011hm} 
  B.~M.~Fisher {\it et al.},
  Nucl.\ Instrum.\ Meth.\ A {\bf 646}, (2011) 126
  
\bibitem{SNOLi}
M.~Yeh, 
Advances in Neutrino Technology, Lake Tahoe, CA, May 12, 2013. 

  
\bibitem{Verkerke:2003ir} 
  W.~Verkerke and D.~P.~Kirkby,
  eConf C {\bf 0303241}, MOLT007 (2003)
  [physics/0306116].
    
\bibitem{Djermouni} 
B.~Djermouni and H.~J.~Ache,
J.\ Phys.\ Chem.\  {\bf 82}, (1978) 2378.  
\end{thebibliography}
\end{document}